\documentclass[11pt, letterpaper]{article}

\usepackage{indentfirst}
\usepackage{verbatim}
\usepackage[dvips]{graphicx}
\usepackage{amsmath}
\usepackage{amsthm}
\usepackage{amssymb}
\usepackage{bm}

\usepackage{graphicx}
\usepackage{wrapfig}
\usepackage{verbatim}
\usepackage{helvet}
\usepackage{float}
\usepackage{calc}
\usepackage{xspace}
\usepackage{pifont}
\usepackage{multirow}
\usepackage[normalem]{ulem}
\usepackage{hyperref}
\usepackage{color}
\usepackage{xcolor}
\usepackage{lmodern}
\usepackage{tikz}
\usetikzlibrary{fit}
\usepackage{paralist}
\usepackage{tcolorbox}
\usepackage{wrapfig}
\usepackage{authblk}
\usepackage[top=1in,bottom=1in,left=1in,right=1in,foot=0.5in,]{geometry}
\usepackage{multicol}
\usepackage{placeins}
\usepackage{csquotes}
\usepackage{subfig}
\usepackage{lineno}
\usepackage[backend=bibtex,style=numeric,sorting=none,maxbibnames=10,minbibnames=10]{biblatex}
\DeclareFieldFormat{annotation}{#1}
\renewbibmacro*{finentry}{%
  \finentry%
  \iffieldundef{annotation}{}{%
    \par\nobreak\vspace{1pt}{\small\printfield{annotation}}}}

\usepackage[font=footnotesize,labelfont=bf]{caption}
\usepackage{tikz}
\usetikzlibrary{backgrounds}
\usepackage{tikzscale}
\usetikzlibrary{circuits.logic.US,arrows,positioning,shapes.geometric}

\usepackage{placeins}

\usepackage{color}
\usepackage{xcolor}

\makeatletter
\patchcmd{\@maketitle}{\LARGE \@title}{\fontsize{16}{19.2}\selectfont\@title}{}{}
\makeatother

\title{
The Connectome and the Quest for the Functional Logic of the \textit{Drosophila} Early Olfactory System}
\author[1$\dagger$*]{Aurel A. Lazar}
\author[2$\dagger$]{Yiyin Zhou}
\affil[1]{Department of Electrical Engineering, Columbia University, New York, NY 10027, USA}
\affil[2]{Department of Computer and Information Science, Fordham University, New York, NY 10023, USA}
\affil[$\dagger$]{Authors' names are listed in alphabetical order.}
\affil[*]{Corresponding author: aurel@ee.columbia.edu}
\begin{document}

%Department of Electrical Engineering, Columbia University, New York, NY 10027
% \date{}
\maketitle

\begin{abstract}

% 250 words max.

In recent decades, the early olfactory system (EOS) of the fruit fly
has become a leading model for studying olfactory processing and associative memory, 
owing in part to a well-characterized feedforward pathway that feeds the 
processes underlying associative memory and by
examining the role played by
 a handful of neurons and synapses. 
The recent completion of dense electron-microscopy connectomes
provides high quality visualizations of every cell type, neuron, and synapse along the early olfactory pathway. 
Yet a wiring diagram, however complete, does not by itself reveal
the functional logic of a neural circuit. 
Reviewing the EOS connectome and synaptome datasets of the past
fifteen years, we note that the feedforward pathway is embedded in dense
local feedback circuits 
of large scale multi-input multi-output neurons.
A systematic understanding of feedback loop abstractions,
and their capacity to govern the input/output transformations at each neuropil stage,
is the underlying foundation of the functional logic of the early olfactory circuits.
In addition, we argue that a
quantitative account of the functional logic requires an explicit model of the odorants present in the natural environment.
Consisting of odorant objects, such a model defines the semantics and syntax
of olfactory information processing, and calls for new distance measures for classifying the odorant semantics in support of associative memory operations.
Furthermore, odor information processing must abide by causality, treating the circuit
as a real-time, stage-by-stage cascade of giant local feedback loops.


\end{abstract}

\newpage

\section{The Quest for Modeling the Morphology of the \textit{Drosophila} EOS}

Fruit flies, like many other insects and vertebrates, possess extraordinary abilities to
distinguish and recognize different smells thanks to their
hard-wired, innate
brain circuits that have been preserved in almost all species \cite{AY05} and evolved according to the environmental niche that they live in.
These brain circuits are genetically determined, and yet the flies detect with ease
odorant changes in their living environments and continuously adjust their behavior.

\begin{figure}[b!]
\centering
\includegraphics[width=0.7\textwidth]{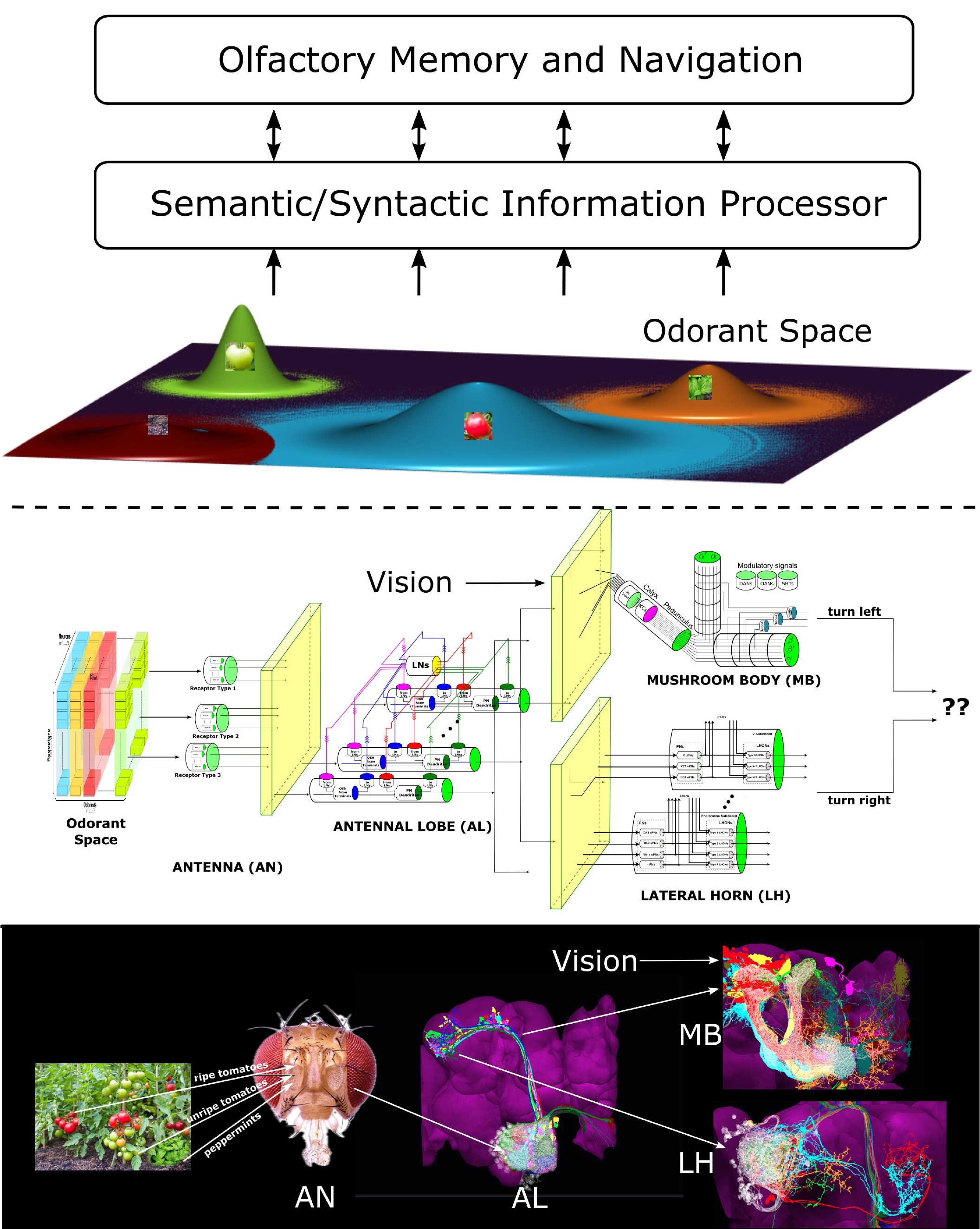}
\caption{The early olfactory system of the fruit fly
supports a semantic and syntactic information processor for navigation
in the odorant space. 
(bottom) The EOS pathway in the \textit{Drosophila}.
Odorants sensed and transduced
by OSNs in the Antenna (AN).
OSN axons forward spike trains
to the Antennal Lobe (AL). Outputs
of the AL then project to both the Mushroom Body (MB) and
the Lateral Horn (LH).
(middle) A pathway abstraction of the EOS circuit
that extracts the canonical circuits and connectivity features
within each neuropil.
(top) To support navigating in an olfactory environment 
a semantics/syntactic information processor 
extracts olfactory objects distributed across the odorant space.
}
\label{fig:osp}
\end{figure}

Few neural circuits are as seeminly simple yet elegant as the early olfactory system (EOS)
of \textit{Drosophila melanogaster}, whose feedforward pathway was identified
in detail well before dense connectome datasets became
available. Figure~\ref{fig:osp} (bottom) from left to right traces this
pathway stage by stage. Odorants are first detected in the Antenna (AN) by olfactory
sensory neurons (OSNs).
Each OSN is typically expressing a single type of odorant
receptor \cite{CAD05,LV08}. The OSN axons project into the antennal lobe (AL),
where all OSNs expressing the same receptor type converge onto a common glomerulus, and thereby receptor identity is mapped onto glomerular identity \cite{SLB90,VWA00};
within each glomerulus the OSNs relay onto the uniglomerular projection neurons
(PNs). The PNs in turn carry this glomerular representation along parallel tracts
to the two higher olfactory centers \cite{MJK02,WWA02}: the Mushroom Body (MB)
Calyx, where a small population of $\sim50$ PN types expands onto
$\sim2000$ Kenyon cells (KCs) through a largely random
convergence \cite{CRA13,TBL08}, and, in parallel, the lateral horn (LH), where the
PN targets are more stereotyped \cite{JPC07}. Within the MB, the KC axons are
divided into 3 bundles and 
converge onto a set of 15 compartments where a small set of MB output neurons (MBONs) read from KC outputs \cite{AHY14}. 
In addition, dopaminergic neurons (DANs) provide additional inputs,
representing, for example, punishment~\cite{CCR09}, reward~\cite{LPT12},
internal states~\cite{KDW09}, and context~\cite{CMR15}, to the MB
compartments, to be associated with odorant inputs carried by the
KCs~\cite{AHY14,ASR14}.

Historically, this feedforward picture has carried most of our understanding of how
the EOS works.
Yet this simple view cannot account for the marked changes in the
population responses from one stage to the next \cite{SDR17,SOI08,IBW07}.
Substantial processing therefore takes place
locally, within each neuropil. Understanding what is computed, and how,
requires the detailed local circuitry, including the interneurons and feedback connections
that reshape the information processed at each neuropil. 

Over the past fifteen years, anatomy marked by genetic labeling and connectomics has begun to supply exactly this missing
ingredient \cite{TEI12,LLM20}. Dense, synapse-scale reconstruction now provides, for each stage of
the EOS, a near-complete catalog of its neurons together with a quantitative map of
how they are wired, including the local neurons, some at remarkable scale and
complexity, that carry out much of the local processing load.

This sets up the two questions that will be echoed throughout the rest of this
review: first, how can such a detailed connectome help us
understand the functional logic of the EOS; and second, beyond the
wiring itself, what else is needed for achieving an understanding of the functional logic of the EOS?

In general, the workflow for exploring the detailed connectome
starts by modeling
the morphology of the EOS circuit (Figure~\ref{fig:osp} bottom row)
and is followed by creating feedback pathway abstractions that extract
processing units and their
modes of interactions (Figure~\ref{fig:osp} middle row).
Using a model of the space of odorants
a Semantic/Syntactic Information Processor provides the high level
functional logic support for Olfactory Memory and Navigation(Figure~\ref{fig:osp} top row)
\cite{LLTZ21}.

In addition, we argue here that the understanding of how EOS work in the
feedforward view has remained largely \textit{qualitative} and anchored in behavior.
By doing so, the concept of \textit{causality}, \textit{i.e.}, the
response of a circuit is only determined by past and present input values,
has rarely been investigated and correlation methods have largely been favored.
Understanding the functional logic of a brain circuit, however,
requires a time-dependent characterization
of the underlying circuit mechanism.
Causality can no longer be disregarded.
Investigating how, for example,
associative learning works in the MB compartments,
requires the time-domain characterization of
 the inputs to the compartments.
For the EOS that requires investigating the I/O stage
by stage, i.e., from the left to the right in Figure~\ref{fig:osp} bottom,
till we reach the input of MB compartments.
In other words, the stage-by-stage causality is a key challenge towards understanding the functional logic of
EOS: how is the representation of odor information progressively transformed across different
neuropil stages (Figure~\ref{fig:osp} middle row) into formats that lead to
classification of the odorant semantics and support memory and navigation \cite{YOH23}.

\section{Connectome/Synaptome Cell Types, and Connectivity Pathways and Feedback Abstractions}

The first brain-wide surveys relied on light microscopy, registering
thousands of genetically labeled single neurons into a common template to catalog the
fly's cell types and their coarse projection patterns~\cite{CCH11}.
Volume electron
microscopy (EM) then made synapse-resolution reconstruction possible, first
in the lamina and medulla of the adult optic lobe \cite{RVM11,TBL13}, then
smaller larval nervous system~\cite{OSF15,BKC16,WPB23} and subsequently across the adult
brain: the complete adult female brain volume (FAFB)~\cite{ZLP18}, the densely
reconstructed central brain of the hemibrain~\cite{SXJ20}, and, most recently, the
complete adult female brain reconstructed in FlyWire~\cite{DMS24,SYB24} and the
male central nervous system~\cite{BBC25}.

These datasets deliver three things that were
previously out of reach. First, they provide a nearly complete parts list: an
inventory of every neuron and its assigned cell type in each neuropil~\cite{SYB24}. Second, they recover the full three-dimensional morphology of each neuron,
resolving the fine dendritic and axonal arbors on which synapses are placed. Third,
and most consequentially, they supply a \textit{synaptome}: the location, count,
and partners of the individual synapses, now augmented by machine-learning
predictions of each synapse's neurotransmitter identity~\cite{EBC24}, so that
connections can be assigned as excitatory or inhibitory.

The early olfactory system has been an early and recurring target of these
efforts. 
The key argument we develop in this review is that, across the neuropils of the
EOS, the connectome does not merely confirm the expected feedforward relays.
It reveals that each processing stage is embedded in, and shaped by, a dense 
set of \textit{feedback} circuits.

\begin{figure}
\centering
\includegraphics[width=1\textwidth]{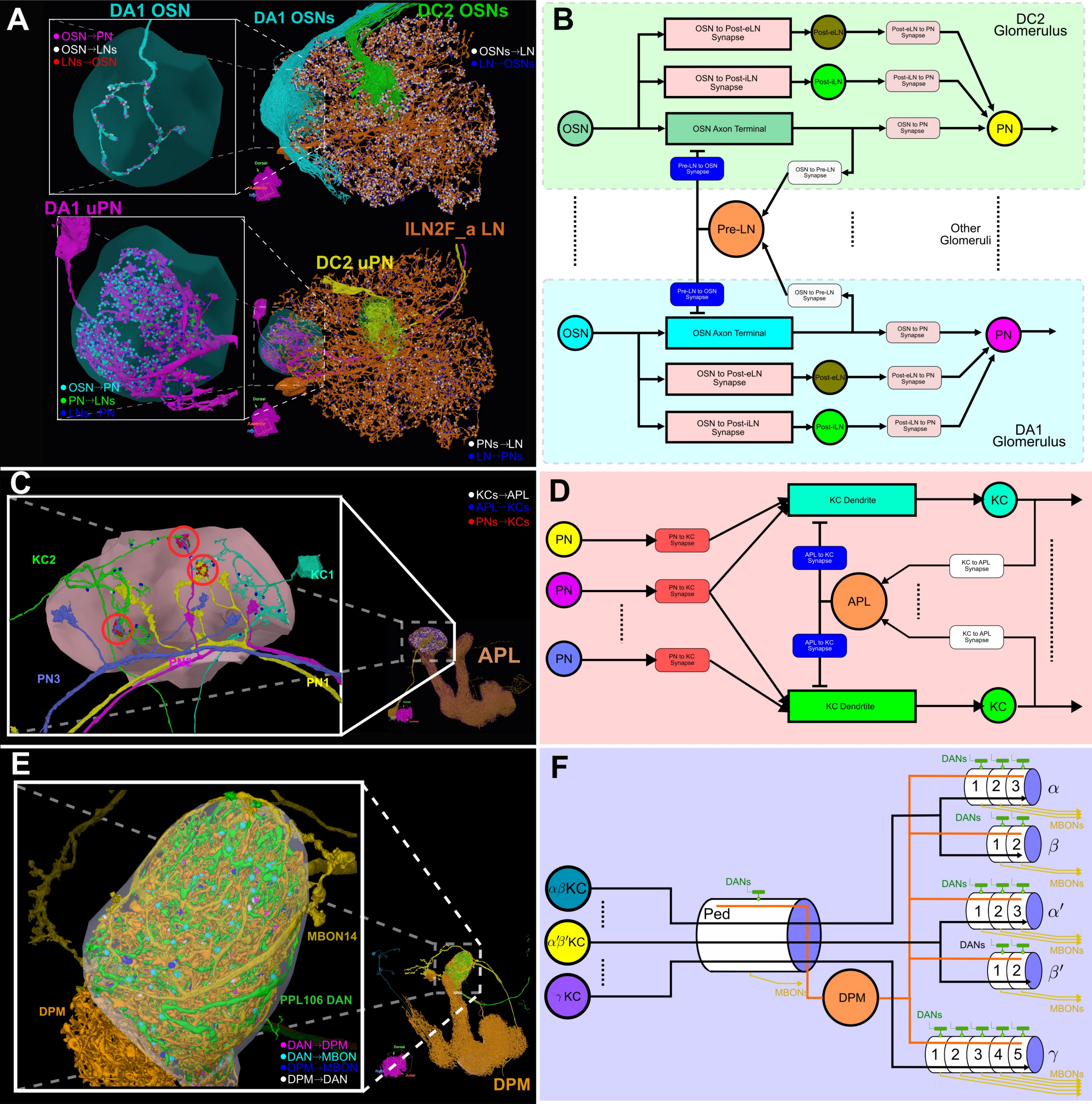}
\caption{
\textbf{(A)} Cell types and connectivity in the AL. Top: (orange)
A full LN (lLN2F\_a) that innervates all glomeruli and forms
a global feedback loop with OSNs.
OSNs that project into the DA1 and DC2 glomeruli are shown
in cyan and green, respectively. In the inset, an OSN
that projects into the DA1 glomerulus is shown, along
with the outline of the glomerulus.
Bottom: The same LN's (in orange) interaction with uPNs.
Two uPNs are shown, a DA1 uPN in magenta and a DC2 uPN
in yellow.
In the inset, a DA1 uPN is shown in magenta, inside the
outlined the DA1 glomerulus. 
\textbf{(B)} Feedback abstraction of the AL. Adopted from \cite{LLYZ24}.
\textbf{(C)} Cell types and connectivity in the MB Calyx. APL neuron is
shown in orange in the lower right. Inset shows a closeup of the Calyx, where
3 PNs and 2 KCs are depicted, as labeled.
\textbf{(D)} Feedback abstraction of the MB Calyx. Adopted from \cite{LLYZ24}.
\textbf{(E)} Cell types and connectivity in a MB compartment. 
The DPM neuron is shown in orange in the bottom right.
The inset shows
a closeup of the $\alpha3$ compartment, involving the DPM neuron
(orange), an MBON14 neuron (yellow) and a PPL106 DAN neuron
(green).
\textbf{(F)} Feedback abstraction of the MB compartments.
Data from the Hemibrain dataset \cite{SXJ20} visualized in NeuroNLP \cite{LLTZ21}.
 }
\label{fig:feedback}
\end{figure}

In the Antennal Lobe, in addition to OSNs and PNs, a diverse population
of local neurons knits
the glomeruli together \cite{SRW10,TEI12,BKC16,SBS21,LTZ22}.
In the Hemibrain dataset over 70 types of different local
neurons exist in the adult AL.
Figure~\ref{fig:feedback}A shows one such instance of a local neuron
that arborizes in all glomeruli. In each glomerulus,
it receives inputs from both OSN axons (top inset) and PN dendrites (bottom inset),
and provides feedback to these.
The same local neuron also interacts with many
other types of LNs throughout the glomeruli (not shown).
Recent experiments have shown 
that the LNs have diverse physiological properties, with some nonspiking 
ones interacting locally within each glomerulus, and some others
operating  globally \cite{BDW23}.

The feedback abstraction of the Antennal Lobe is 
summarized in Figure~\ref{fig:feedback}B. In each glomerulus, an excitatory
Post-eLNs
and an inhibitory Post-iLN act within the glomerulus
and a Pre-LN sends feedback to the axon terminal of the OSNs.

In the Mushroom Body Calyx, the single giant APL
neuron collects input from the axon terminals of PNs and from the dendrites of 
Kenyon cells, and feeds back onto both \cite{LLM20},
as shown in Figure~\ref{fig:feedback}C.
The feedback abstraction of the Calyx is shown in Figure~\ref{fig:feedback}D,
which highlights the expansion recoding \cite{MST20}
from PNs to KCs as well as the global feedback facilitated by the APL neuron.

In each Mushroom Body compartment (with the $\alpha3$-compartment shown in
Figure~\ref{fig:feedback}E), the locally connected KC axon, DAN inputs and
MBON dendrites form feedback loops across all compartments within
a tightly recurrent motif consisting of the APL neuron
and another giant DPM neuron \cite{AHY14,LLM20}.
In addition, outside of MB, the DANs may receive inputs from MBONs that
target the same and/or a different compartment \cite{EFW20,LLM20}.
Therefore, the DAN inputs
coming into each compartment can be also considered
to be part of feedback loops.

In Figure~\ref{fig:feedback}E we highlight the DPM neuron
as the key component of the feedback loop
abstraction in the MB compartments.
Three bundles of KCs, namely, the $\alpha\beta$KCs, $\alpha'\beta'$KCs and
the $\gamma$KCs, all pass through the Pedunculus and then each project to 5 compartments in, respectively $\alpha$ and $\beta$ lobes, $\alpha'$ and $\beta'$ lobes, and the $\gamma$ lobe. At least one type of DANs also project into each
compartment where one or more MBONs carry the outputs.

In every case the same
design recurs: a thin feedforward backbone wrapped in a much heavier layer of
feedback, often times involving some giant neurons. It is this feedback architecture, rather than the feedforward backbone it
modulates, that the connectome uniquely exposes, and we argue that it should be the primary target of any account of the functional logic of the EOS.

\section{The Quest for the Functional Logic of Feedback Circuits}

And yet a catalog of who connects to whom, however
complete, has not by itself told us what the circuit computes.
What is extracted in each stage from the odor information waveforms?
What roles do these large scale multi-input multi-output (MIMO) feedback
neurons play remains surprisingly hard to pin down, particularly in
system neuroscience experiments due to the extraordinarily large number
of input/outputs they connect to.

\FloatBarrier

\subsection{The Odorant Space Dictates the Functional Logic of the Antenna Circuit}

To understand what odorant information
is extracted at each stage in the EOS,
we must first define what olfactory information is.
That demands carefully modeling objects in the space of odorants and their
interaction with the odorant receptors uniquely expressed by the OSNs.
This essential step in the
quest for determining the functional logic of olfactory brain
circuits has often and largely been neglected.

A number of approaches have addressed the
relationship between the chemical structure
and the response of the OSNs to odorants.
Given the vastly large dimensionality
of  the molecular chemistry of odorants, the goal
was to find a low dimensional odorant model that
exhibits the key odorant features.
This approach has recently attracted more
attention thanks to recent advances in artificial neural networks, including
predictors based on convolutional autoencoders \cite{TKS19} and graph neural
networks \cite{QWS23}.

While these approaches typically inquire how olfactory receptors respond
to odorants, they rarely describe how molecular information
can help understanding the processing of odor information
waveforms in later stages of the EOS. 
In other words, just as visual objects in early vision are not directly modeled using
electromagnetic waves,
chemical composition may not be the most appropriate way of defining odorant objects.

In another approach, the objects in the odorant space are 
explicitly and separately modeled by the semantic and syntactic characterization of
pure odorants \cite{LLY23}.
The semantics is described by the interaction between odorants
and olfactory receptors as a pair of tensors \cite{LAY18, LAY20}
determined by the ON-OFF process of binding and dissociation of the odorant
with the OSN receptors.
The syntax is defined as the odorant concentration waveform.
Both the binding/dissociation rate and odorant concentration
waveform jointly determine the outcome of the odorant transduction process \cite{LAY20}.
Therefore, the multiplicative coupling between the odorant semantics and
the odorant syntax determine the functional logic of the antenna circuit.

More importantly, the model of objects in the odorant space also sets the 
key objective of early olfactory processing: to extract
the identity, or odorant semantics, 
from the multiplicatively coupled odorant concentration, or syntax.
On the one hand, it is important to recognize that the concentration
waveform (syntax) is a function of time, and flies navigate in a
dynamic environment where odors come and go, and their concentration
gradient could be critical to search and find odorant
objects (see also Figure~\ref{fig:osp} top).
As the extraction of odorant identity takes place
in time and it is thereby time-dependent, it must adhere to causality. 
Nonetheless, most models of the odorant space in the literature have neglected the
temporal aspect of olfactory processing. 
We note that, the odorant concentration waveform (syntax) represents
the traditional information paradigm in the Shannon sense and
lacks the notion of ``meaning" or semantics.
Together, odorant identity modeled as ON-OFF semantics calls
for semantic information processing in the time domain.

A similar approach for modeling odorants uses the affinity
between odorants with each receptor
to construct a ``primacy hull" \cite{GSR24}, where the identity
of odorant is determined by the set of OSNs with the strongest responses.
However, this model of odorants
lacks time dependency, i.e., the odorant syntax information
component.

\subsection{Antennal Lobe Extracts Odorant Identity and ON-OFF Timing Events}

Many models of olfactory processing consider PNs to carry the same information as OSNs.
However, this grossly overlooks the role of 
all the LN cell types and the feedback they provide in AL processing.
LNs in the AL play a key role in decoupling the semantic and syntactic information
encoded in the OSN axons.

With the two types of local neurons depicted in Figure~\ref{fig:feedback}B, and the understanding
that the input to the AL is the multiplicatively coupled odorant
semantic and syntactic information, \cite{LLY23}
argues that the two LN type each extracts different features of the odorant.
Two Divisive Normalization Processors (DNPs) each modeling, respectively,
the Post-iLNs and Post-eLNs act locally in each channel and extract
the ON/OFF timing events, while another DNP modeling the Pre-LN
through global feedback extracts the odorant semantics.
The two DNPs reproduces experimentally
recorded PN data \cite{LLY23,KLS15}.
Therefore, the functional logic of the AL is that of an ON-OFF odorant
object identity recovery processor, determining the identity of 
the odorant during the time it is presented.

\subsection{Mushroom Body Calyx Generates the Input Code for MB Associative Learning}

The outputs of the Mushroom Body Calyx are the KC axons. They supply
odorant semantics directly into the MB associative learning circuits.
Therefore, KCs must carry a code that is ready for memory storage and
access, and has enough discrimination power to tell apart different odorants. 

The quest for the functional logic of the Calyx has been focused
on the two pathway features abstracted in Figure~\ref{fig:feedback}D.
Why is there an expansion from $\sim50$ types of PNs to $\sim2000$ KCs via
a largely random connectivity?
What is the role of the APL feedback neuron in conjunction with this
expansion? 
Several models provided alternate interpretations of the functional logic
of the Calyx.

\cite{DSN17} argues that random projections from PNs to KCs with 
winner-take-all KCs output facilitated by APL
feedback lead to a locality-sensitive hashing function.
The hashing function efficiently
assigns similar KC activity to similar odorants and
thereby, subsequent
associations of one odorant can be generalized to similar odorants.
Here, the assumption is that the firing rate
of PN inputs to the Calyx can be modeled with an exponential distribution.

Recently, a different model considers the
PNs to KCs connectivity matrix,
along with the KCs to MBONs connectivity matrix, as part
of a random measurement (compressed sensing) matrix.
This allows the representation of odorant waveforms
through compressed sensing at the MBON level \cite{CKH24}. 

It has been experimentally demonstrated that the
APL neuron normalizes odor-evoked
responses in individual KCs \cite{PDY21}.
However, how does this process relate\
to the odor information flow generated by the population of KCs?
By modeling the global feedback
as a differential Divisive Normalization Processing circuit, the Calyx transforms the representation of
ON-OFF odorant semantics at the output of the AL into a marked first spike sequence code \cite{LLYZ24,LZ26}.
This code reflects the amplitude \textit{ranking} that
drives the KCs in the time domain.

\begin{figure}[t!]
\centering
\includegraphics[width=1\textwidth]{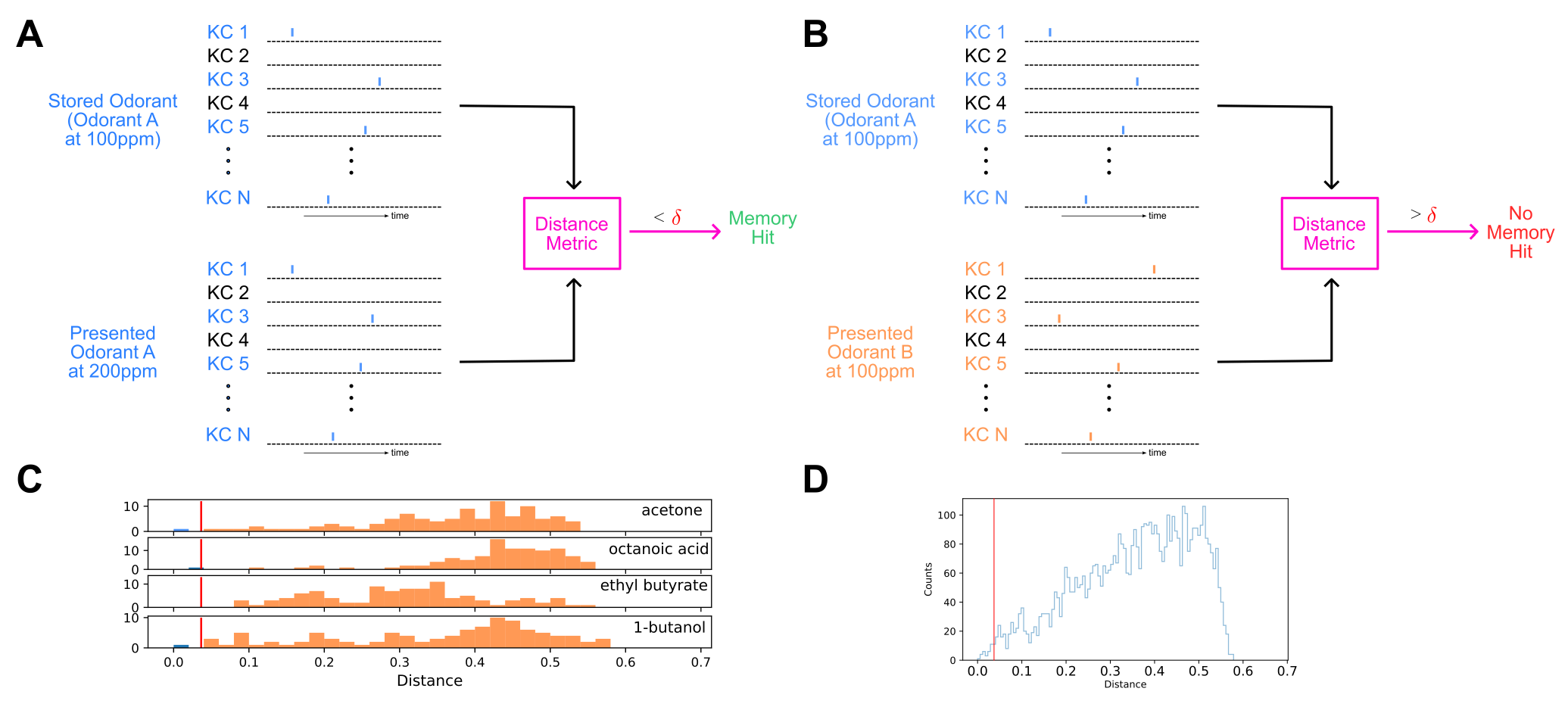}
\caption{Semantic information represented in the KCs 
supports high capacity associative memory operations.
\textbf{(A)} (top left) KC spiking representation of the odorant
identity (semantics) that was already stored
in the MB (Odorant A at 100 ppm).
(bottom left) KC spiking representation of the same odorant at 200 ppm.
A low distance measure \cite{LZ26}, achieved largely regardless
of the amplitude of the concentration waveform, leads to a memory hit.
\textbf{(B)} KC representations of \textit{different} odorant identities
(semantics) leads to a greater rank-based distance measure
for odorant semantics (see text), signaling that the odorant
presented has not been stored. Here, even if the odorant input 
evoke response in the same set of KCs as the stored odorant activated,
the ordering (mark) of the spikes differ substantially to indicate a miss.
\textbf{(C-D)} The marked first spike sequence code is highly effective
in differentiating different odorants while preserving a low distance
between the same odorant presented at different concentrations.
\textbf{(C)} Each row is a histogram of the distances of each of 103 odorants
\cite{HC06} from the labeled odorant.
The red vertical line is the threshold $\delta$ set by the maximum distance 
between different concentration amplitudes of the same odorant across all 104 odorants simulated.
\textbf{(D)} Histogram of all pair-wise distances between pairs of 104
odorants. The vertical line indicates the threshold as in \textbf{(C)}.
Only a small fraction will cause a memory hit error
(left to the red vertical line).
}
\label{fig:classification}
\end{figure}

When paired with a variation of the Kemeny-Snell
distance measure \cite{KS62},
classification of the odorant semantics based on the marked first spike
sequence is achieved with high accuracy.
The APL feedback removes the concentration
dependency of the KC outputs, thereby reducing to small
values the distance between the same
odorant at different amplitude concentration levels (Figure~\ref{fig:classification}A).
It also increases the ranking distance between the
marked first spike sequence codes representing different odorant identities (Figure~\ref{fig:classification}B-D),
whereas odorants encoded with traditional binary KC code may still remain substantially overlapping \cite{kennedy19}.
As the natural representation of semantics accessing the associative memory circuit lies in the spike domain, this possibly supports a low complexity rapid
readout of the semantic information at the KC-MBON synapses primed for
associating the odorant semantics with semantic information
carried by the DANs. 

This suggests that even though the PN to KC
 connections are random 
across individual flies, overall relationships among odorants represented
at the output of the Calyx are highly similar across flies \cite{YOH23}.

In addition to olfactory information, some subtypes of KCs receive exclusive
visual inputs \cite{GHL24} (see also Figure~\ref{fig:osp} bottom). This means that visual and olfactory information are
integrated in the Mushroom Body during learning \cite{OOC26}.
Integrating the full syntactic detail of each modality would be computationally prohibitive at this stage; it is far
more economical to combine the modalities only after each has been reduced to its
\textit{semantic} content. The concentration-invariant first spike sequence code
generated in the Calyx is precisely such a representation.

\subsection{Models of Associative Learning in the Mushroom Body}
\label{sec:mbc}

The goal of \textit{associative learning} in the MB is to
\textit{assign a semantic value} to olfactory stimuli in real-time \cite{BAW20}.
Recent advances have established a clear framework for \textit{Drosophila}
associative learning operations \cite{MST20}.

As shown in Figure~\ref{fig:feedback}F, the MB is divided into 
multiple compartments in which specific reward inputs target
specific output neurons. In each compartment, KCs, MBONs and DANs
form distinct tripartite synapses, suggesting a more complex flow of
information between them.
This has been the main focus of modeling the learning rule 
that underlies associative memory in the MB.
Modelers, however, have not yet converged to an agreement \cite{webb24}.
Where the models differ is in what triggers the plasticity
at KC$\rightarrow$MBON synapses.

Webb's survey \cite{webb24} distinguishes three broad families. In the first,
plasticity is Hebbian: the synapse
is modified when KC and MBON activity coincide. In the second, plasticity is
neuromodulatory: the change is gated by a dopaminergic neuron (DAN) that
targets the synapse and delivers a teaching signal, interpreted variously as a
reward or as a prediction error.
In the third, the two are combined, so that synaptic change
requires both coincident KC activity and a DAN signal.
A recurring assumption across
many of these models, inherited from top-down theories of associative learning, is
that the circuit's core function is to minimize prediction error; Webb argues that a
more bottom-up approach may instead reveal a richer algorithmic capacity in this
neuropil \cite{webb24}, which may explain why MBON to DAN feedback loops are quite common.

However, as we have shown in Figure~\ref{fig:feedback}F, 
this seems to be only part of the story. 
We would like to point out several critical points 
that most models so far neglected.

First, the contribution of the giant DPM and APL neurons is
rarely incorporated into the learning model.
Even though experiments suggested that disturbing DPM output
does not affect the conditioning, it is critical in
the consolidation of memory after association \cite{YKS05}.
Recent evidence
showed that the DPM neuron also played a central
role in defining the time window
during which association can be formed \cite{ZLZ23}.

Second, most of the models do not take into account
what information is presented at the input of the MB
compartments and how it is represented.
As a result, it is not clear whether the MB compartment circuit 
is able to distinguish odorant semantics,
how much computational power is required to sort out
this part, or the capacity of memory that the circuit can offer.
 As we have argued so far,
without a characterization of causality, the understanding
of the local circuitry can only be, at best, partial.
The marked first spike sequence code and 
other existing codes \cite{LZ26,DSN17,kennedy19} would 
impose different requirements on the associative learning circuit.

Third, associative learning in the MB is also time dependent.
It has been shown that presenting the DAN input at different times
with respect to odorant presentation can produce different, and sometimes
completely opposite, effects \cite{THG04,HGC19}.
Therefore, it is important for the ON-OFF timing of the odorant semantics to be extracted and represented. Testing models of
associative learning therefore requires to take into account these codes, 
such as the marked first spike sequence code.

\section{Conclusions}

The connectome of the EOS will continue to play a central role in the quest for its
functional logic. The local processing that
transforms the signal at each stage is carried out not by simple relays but by dense
\textit{feedback} circuits, built around MIMO neurons.
Since 
directly probing such high-fan-in, high-fan-out neurons is beyond the reach of the
current experimental landscape, computational models that take these feedback
circuits as their primary object are not merely useful but necessary.

Feedback circuitry, however, is only one of the many ingredients that a quantitative
account of its functional logic demands. What a circuit computes is defined
relative to the structure of its inputs, and for the EOS this means an explicit
model of odorant objects in the odorant space characterized by
odorant semantics and syntax.
Without such a model one cannot fully substantiate what a feedback circuit is
actually computing, nor how its operation depends on the inputs it receives.
In addition, neural circuits operate in real-time and in cascade: the input to each neuropil is the output of the one before it, so the input to a neuropil
stage cannot simply
be assumed but must be generated by the preceding neuropil
stages of the model.
Causality needs to be established in real-time to turn a static wiring diagram
into a quantitative account of what the functional logic of the circuit
might be.

Finally, what we have discussed in this review does not only apply to olfaction. The same feedback-dominated
architecture, and the same need to specify both the input space and the causal
cascade, recur throughout the sensory systems of fly brain.
In vision, for example, the large
local neurons of the Medulla and Lobula Plate carry out comparable local
computation. A quantitative understanding of the functional logic of the
\textit{Drosophila} EOS, and of the circuits beyond it, will come only from models that hold these keys together.
Moreover, the results for the EOS of the fruit fly presented in this review
may shine a light on the principles of olfactory processing in other animals, 
including primates and humans where it may still prove difficult to obtain a connectome.

\subsection*{Acknowledgement}
The research reported here was supported, in part, by the National Science Foundation under grant \#2400687.

\subsection*{Declaration of Interest statement}
The authors declare no competing interest.

\newpage

\printbibliography

@inbook{KS62,
	address = {Cambridge, MA},
	author = {Kemeny, J.G. and Snell, J.L.},
	chapter = {2. Preference rankings an axiomatic approach},
	pages = {9-23},
	publisher = {The MIT Press},
	title = {Mathematical Models in the Social Sciences},
	year = {1962}}

@article{LLTZ21,
	article_type = {journal},
	author = {Lazar, Aurel A. and Liu, Tingkai and Turkcan, Mehmet Kerem and Zhou, Yiyin},
	citation = {eLife 2021;10:e62362},
	doi = {10.7554/eLife.62362},
	editor = {Bhalla, Upinder Singh and Calabrese, Ronald L and Gleeson, Padraig},
	issn = {2050-084X},
	journal = {eLife},
	month = {feb},
	pages = {e62362},
	pub_date = {2021-02-22},
	publisher = {eLife Sciences Publications, Ltd},
	title = {Accelerating with FlyBrainLab the discovery of the functional logic of the \textit{Drosophila} brain in the connectomic and synaptomic era},
	url = {https://doi.org/10.7554/eLife.62362},
	volume = 10,
	year = 2021}

@article{AY05,
	author = {Ache, Barry W. and Young, Janet M.},
	date = {2005/11/03},
	doi = {10.1016/j.neuron.2005.10.022},
	isbn = {0896-6273},
	journal = {Neuron},
	journal1 = {Neuron},
	month = {2021/12/13},
	number = {3},
	pages = {417--430},
	publisher = {Elsevier},
	title = {Olfaction: Diverse Species, Conserved Principles},
	type = {doi: 10.1016/j.neuron.2005.10.022},
	url = {https://doi.org/10.1016/j.neuron.2005.10.022},
	volume = {48},
	year = {2005},
	year1 = {2005}}

@article{IBW07,
	author = {Olsen, Shawn R. and Bhandawat, Vikas and Wilson, Rachel I.},
	date = {2007/04/05},
	doi = {10.1016/j.neuron.2007.03.010},
	isbn = {0896-6273},
	journal = {Neuron},
	journal1 = {Neuron},
	month = {2026/08/16},
	n2 = {Each odorant receptor gene defines a unique type of olfactory receptor neuron (ORN) and a corresponding type of second-order neuron. Because each odor can activate multiple ORN types, information must ultimately be integrated across these processing channels to form a unified percept. Here, we show that, in Drosophila, integration begins at the level of second-order projection neurons (PNs). We genetically silence all the ORNs that normally express a particular odorant receptor and find that PNs postsynaptic to the silent glomerulus receive substantial lateral excitatory input from other glomeruli. Genetically confining odor-evoked ORN input to just one glomerulus reveals that most PNs postsynaptic to other glomeruli receive indirect excitatory input from the single ORN type that is active. Lateral connections between identified glomeruli vary in strength, and this pattern of connections is stereotyped across flies. Thus, a dense network of lateral connections distributes odor-evoked excitation between channels in the first brain region of the olfactory processing stream.},
	number = {1},
	pages = {89--103},
	publisher = {Elsevier},
	title = {Excitatory Interactions between Olfactory Processing Channels in the \textit{Drosophila} Antennal Lobe},
	type = {doi: 10.1016/j.neuron.2007.03.010},
	url = {https://doi.org/10.1016/j.neuron.2007.03.010},
	volume = {54},
	year = {2007},
	year1 = {2007}}

@article{SOI08,
	author = {Silbering, Ana F. and Okada, Ryuichi and Ito, Kei and Galizia, C. Giovanni},
	doi = {10.1523/JNEUROSCI.2973-08.2008},
	eprint = {https://www.jneurosci.org/content/28/49/13075.full.pdf},
	issn = {0270-6474},
	journal = {Journal of Neuroscience},
	number = {49},
	pages = {13075--13087},
	publisher = {Society for Neuroscience},
	title = {Olfactory Information Processing in the Drosophila Antennal Lobe: Anything Goes?},
	url = {https://www.jneurosci.org/content/28/49/13075},
	volume = {28},
	year = {2008}}

@article{SDR17,
	author = {Seki, Yoichi and Dweck, Hany K. M. and Rybak, J{\"u}rgen and Wicher, Dieter and Sachse, Silke and Hansson, Bill S.},
	date = {2017/06/30},
	doi = {10.1186/s12915-017-0389-z},
	id = {Seki2017},
	isbn = {1741-7007},
	journal = {BMC Biology},
	number = {1},
	pages = {56},
	title = {Olfactory coding from the periphery to higher brain centers in the Drosophila brain},
	url = {https://doi.org/10.1186/s12915-017-0389-z},
	volume = {15},
	year = {2017}}

@article{HC06,
	author = {Hallem, Elissa a and Carlson, John R},
	doi = {10.1016/j.cell.2006.01.050},
	issn = {0092-8674},
	journal = {Cell},
	month = apr,
	number = {1},
	pages = {143--60},
	pmid = {16615896},
	title = {Coding of odors by a receptor repertoire.},
	volume = {125},
	year = {2006}}

@article{BAW20,
	author = {Barbara Webb},
	doi = {10.1126/science.aaz6869},
	eprint = {https://www.science.org/doi/pdf/10.1126/science.aaz6869},
	journal = {Science},
	number = {6488},
	pages = {244-245},
	title = {Robots with insect brains},
	url = {https://www.science.org/doi/abs/10.1126/science.aaz6869},
	volume = {368},
	year = {2020}}

@article{OOC26,
	author = {Okray, Zeynep and Otto, Nils and Cook, Anna A and Talbot, Clifford and Miriyala, Ashwin and Klappenbach, Mart{\'\i}n and Stern, Ciara and Desmond, Kieran and Vargas-Gutierrez, Paola and Waddell, Scott},
	doi = {10.7554/elife.111909.1},
	month = Aug,
	publisher = {eLife Sciences Publications, Ltd},
	title = {Multisensory learning recruits visual neurons into an olfactory memory engram},
	url = {http://dx.doi.org/10.7554/eLife.111909.1},
	year = {2026}}

@article{GHL24,
	author = {Ganguly, Ishani and Heckman, Emily L. and Litwin-Kumar, Ashok and Clowney, E. Josephine and Behnia, Rudy},
	date = {2024/07/07},
	doi = {10.1038/s41467-024-49616-z},
	id = {Ganguly2024},
	isbn = {2041-1723},
	journal = {Nature Communications},
	number = {1},
	pages = {5698},
	title = {Diversity of visual inputs to Kenyon cells of the Drosophila mushroom body},
	url = {https://doi.org/10.1038/s41467-024-49616-z},
	volume = {15},
	year = {2024}}

@article{EFW20,
	author = {Eschbach, Claire and Fushiki, Akira and Winding, Michael and Schneider-Mizell, Casey M. and Shao, Mei and Arruda, Rebecca and Eichler, Katharina and Valdes-Aleman, Javier and Ohyama, Tomoko and Thum, Andreas S. and Gerber, Bertram and Fetter, Richard D. and Truman, James W. and Litwin-Kumar, Ashok and Cardona, Albert and Zlatic, Marta},
	date = {2020/04/01},
	doi = {10.1038/s41593-020-0607-9},
	id = {Eschbach2020},
	isbn = {1546-1726},
	journal = {Nature Neuroscience},
	number = {4},
	pages = {544--555},
	title = {Recurrent architecture for adaptive regulation of learning in the insect brain},
	url = {https://doi.org/10.1038/s41593-020-0607-9},
	volume = {23},
	year = {2020}}

@article{WPB23,
	author = {Michael Winding and Benjamin D. Pedigo and Christopher L. Barnes and Heather G. Patsolic and Youngser Park and Tom Kazimiers and Akira Fushiki and Ingrid V. Andrade and Avinash Khandelwal and Javier Valdes-Aleman and Feng Li and Nadine Randel and Elizabeth Barsotti and Ana Correia and Richard D. Fetter and Volker Hartenstein and Carey E. Priebe and Joshua T. Vogelstein and Albert Cardona and Marta Zlatic},
	doi = {10.1126/science.add9330},
	eprint = {https://www.science.org/doi/pdf/10.1126/science.add9330},
	journal = {Science},
	number = {6636},
	pages = {eadd9330},
	title = {The connectome of an insect brain},
	url = {https://www.science.org/doi/abs/10.1126/science.add9330},
	volume = {379},
	year = {2023}}

@article{ZLZ23,
	author = {Zeng, Jianzhi and Li, Xuelin and Zhang, Renzimo and Lv, Mingyue and Wang, Yipan and Tan, Ke and Xia, Xiju and Wan, Jinxia and Jing, Miao and Zhang, Xiuning and Li, Yu and Yang, Yang and Wang, Liang and Chu, Jun and Li, Yan and Li, Yulong},
	date = {2023/04/05},
	doi = {10.1016/j.neuron.2022.12.034},
	isbn = {0896-6273},
	journal = {Neuron},
	journal1 = {Neuron},
	month = {2026/08/14},
	n2 = {The coincidence between conditioned stimulus (CS) and unconditioned stimulus (US) is essential for associative learning; however, the mechanism regulating the duration of this temporal window remains unclear. Here, we found that serotonin (5-HT) bi-directionally regulates the coincidence time window of olfactory learning in Drosophila and affects synaptic plasticity of Kenyon cells (KCs) in the mushroom body (MB). Utilizing GPCR-activation-based (GRAB) neurotransmitter sensors, we found that KC-released acetylcholine (ACh) activates a serotonergic dorsal paired medial (DPM) neuron, which in turn provides inhibitory feedback to KCs. Physiological stimuli induce spatially heterogeneous 5-HT signals, which proportionally gate the intrinsic coincidence time windows of different MB compartments. Artificially reducing or increasing the DPM neuron-released 5-HT shortens or prolongs the coincidence window, respectively. In a sequential trace conditioning paradigm, this serotonergic neuromodulation helps to bridge the CS-US temporal gap. Altogether, we report a model circuitry for perceiving the temporal coincidence and determining the causal relationship between environmental events.},
	number = {7},
	pages = {1118--1135.e5},
	publisher = {Elsevier},
	title = {Local 5-HT signaling bi-directionally regulates the coincidence time window for associative learning},
	type = {doi: 10.1016/j.neuron.2022.12.034},
	url = {https://doi.org/10.1016/j.neuron.2022.12.034},
	volume = {111},
	year = {2023},
	year1 = {2023}}

@article{YKS05,
	author = {Yu, Dinghui and Keene, Alex C. and Srivatsan, Anjana and Waddell, Scott and Davis, Ronald L.},
	date = {2005/12/02},
	doi = {10.1016/j.cell.2005.09.037},
	isbn = {0092-8674},
	journal = {Cell},
	journal1 = {Cell},
	month = {2026/08/14},
	n2 = {Formation of normal olfactory memory requires the expression of the wild-type amnesiac gene in the dorsal paired medial (DPM) neurons. Imaging the activity in the processes of DPM neurons revealed that the neurons respond when the fly is stimulated with electric shock or with any odor that was tested. Pairing odor and electric-shock stimulation increases odor-evoked calcium signals and synaptic release from DPM neurons. These memory traces form in only one of the two branches of the DPM neuron process. Moreover, trace formation requires the expression of the wild-type amnesiac gene in the DPM neurons. The cellular memory traces first appear at 30 min after conditioning and persist for at least 1 hr, a time window during which DPM neuron synaptic transmission is required for normal memory. DPM neurons are therefore ?odor generalists? and form a delayed, branch-specific, and amnesiac-dependent memory trace that may guide behavior after acquisition.},
	number = {5},
	pages = {945--957},
	publisher = {Elsevier},
	title = {\textit{Drosophila} DPM Neurons Form a Delayed and Branch-Specific Memory Trace after Olfactory Classical Conditioning},
	type = {doi: 10.1016/j.cell.2005.09.037},
	url = {https://doi.org/10.1016/j.cell.2005.09.037},
	volume = {123},
	year = {2005},
	year1 = {2005}}

@article{webb24,
	author = {Webb, Barbara},
	doi = {10.1101/lm.053824.123},
	elocation-id = {a053824},
	eprint = {http://learnmem.cshlp.org/content/31/5/a053824.full.pdf+html},
	journal = {Learning \& Memory},
	number = {5},
	title = {Beyond prediction error: 25 years of modeling the associations formed in the insect mushroom body},
	url = {http://learnmem.cshlp.org/content/31/5/a053824.abstract},
	volume = {31},
	year = {2024}}

@article{LZ26,
	author = {Lazar, Aurel A. and Zhou, Yiyin},
	doi = {10.64898/2026.01.05.697602},
	elocation-id = {2026.01.05.697602},
	eprint = {https://www.biorxiv.org/content/early/2026/01/05/2026.01.05.697602.full.pdf},
	journal = {bioRxiv},
	publisher = {Cold Spring Harbor Laboratory},
	title = {Elements of Olfactory Intelligence in \textit{Drosophila}},
	url = {https://www.biorxiv.org/content/early/2026/01/05/2026.01.05.697602},
	year = {2026}}

@article{MST20,
	author = {Modi, Mehrab N. and Shuai, Yichun and Turner, Glenn C.},
	booktitle = {Annual Review of Neuroscience},
	doi = {10.1146/annurev-neuro-080317-0621333},
	isbn = {0147-006X},
	journal = {Annual Review of Neuroscience},
	journal1 = {Annu. Rev. Neurosci.},
	month = {1},
	n2 = {The Drosophila brain contains a relatively simple circuit for forming Pavlovian associations, yet it achieves many operations common across memory systems. Recent advances have established a clear framework for Drosophila learning and revealed the following key operations: a) pattern separation, whereby dense combinatorial representations of odors are preprocessed to generate highly specific, nonoverlapping odor patterns used for learning; b) convergence, in which sensory information is funneled to a small set of output neurons that guide behavioral actions; c) plasticity, where changing the mapping of sensory input to behavioral output requires a strong reinforcement signal, which is also modulated by internal state and environmental context; and d) modularization, in which a memory consists of multiple parallel traces, which are distinct in stability and flexibility and exist in anatomically well-defined modules within the network. Cross-module interactions allow for higher-order effects where past experience influences future learning. Many of these operations have parallels with processes of memory formation and action selection in more complex brains.},
	number = {1},
	pages = {465--484},
	publisher = {Annual Reviews},
	title = {The \textit{Drosophila} Mushroom Body: From Architecture to Algorithm in a Learning Circuit},
	ty = {JOUR},
	url = {https://doi.org/10.1146/annurev-neuro-080317-0621333},
	volume = {43},
	year = {2020},
	year1 = {2020}}

@inproceedings{TKS19,
	author = {Tran, Ngoc and Kepple, Daniel and Shuvaev, Sergey and Koulakov, Alexei},
	booktitle = {Proceedings of the 36th International Conference on Machine Learning},
	editor = {Chaudhuri, Kamalika and Salakhutdinov, Ruslan},
	month = {09--15 Jun},
	pages = {6305--6314},
	publisher = {PMLR},
	series = {Proceedings of Machine Learning Research},
	title = {{D}eep{N}ose: Using artificial neural networks to represent the space of odorants},
	url = {https://proceedings.mlr.press/v97/tran19b.html},
	volume = {97},
	year = {2019}}

@article{DSN17,
	author = {Sanjoy Dasgupta and Charles F. Stevens and Saket Navlakha},
	doi = {10.1126/science.aam9868},
	eprint = {https://www.science.org/doi/pdf/10.1126/science.aam9868},
	journal = {Science},
	number = {6364},
	pages = {793-796},
	title = {A neural algorithm for a fundamental computing problem},
	url = {https://www.science.org/doi/abs/10.1126/science.aam9868},
	volume = {358},
	year = {2017}}

@article{GSR24,
	author = {Giaffar, Hamza AND Shuvaev, Sergey AND Rinberg, Dmitry AND Koulakov, Alexei A.},
	doi = {10.1371/journal.pcbi.1012379},
	journal = {PLOS Computational Biology},
	month = {09},
	number = {9},
	pages = {1-23},
	publisher = {Public Library of Science},
	title = {The primacy model and the structure of olfactory space},
	url = {https://doi.org/10.1371/journal.pcbi.1012379},
	volume = {20},
	year = {2024}}

@article{QWS23,
	article_type = {journal},
	author = {Qian, Wesley W and Wei, Jennifer N and Sanchez-Lengeling, Benjamin and Lee, Brian K and Luo, Yunan and Vlot, Marnix and Dechering, Koen and Peng, Jian and Gerkin, Richard C and Wiltschko, Alexander B},
	citation = {eLife 2023;12:e82502},
	doi = {10.7554/eLife.82502},
	editor = {Bhalla, Upinder Singh and King, Andrew J},
	issn = {2050-084X},
	journal = {eLife},
	month = {may},
	pages = {e82502},
	pub_date = {2023-05-02},
	publisher = {eLife Sciences Publications, Ltd},
	title = {Metabolic activity organizes olfactory representations},
	url = {https://doi.org/10.7554/eLife.82502},
	volume = 12,
	year = 2023}

@article{YOH23,
	author = {Yang, Jie-Yoon and O{\textquoteright}Connell, Thomas F. and Hsu, Wei-Mien M. and Bauer, Matthew S. and Dylla, Kristina V. and Sharpee, Tatyana O. and Hong, Elizabeth J.},
	doi = {10.1101/2023.02.15.528627},
	elocation-id = {2023.02.15.528627},
	eprint = {https://www.biorxiv.org/content/early/2023/02/16/2023.02.15.528627.full.pdf},
	journal = {bioRxiv},
	publisher = {Cold Spring Harbor Laboratory},
	title = {Restructuring of olfactory representations in the fly brain around odor relationships in natural sources},
	url = {https://www.biorxiv.org/content/early/2023/02/16/2023.02.15.528627},
	year = {2023}}

@article{kennedy19,
	author = {Kennedy, Ann},
	doi = {10.1101/783191},
	elocation-id = {783191},
	eprint = {https://www.biorxiv.org/content/early/2019/09/26/783191.full.pdf},
	journal = {bioRxiv},
	publisher = {Cold Spring Harbor Laboratory},
	title = {Learning with naturalistic odor representations in a dynamic model of the Drosophila olfactory system},
	url = {https://www.biorxiv.org/content/early/2019/09/26/783191},
	year = {2019}}

@article{PDY21,
	article_type = {journal},
	author = {Prisco, Luigi and Deimel, Stephan Hubertus and Yeliseyeva, Hanna and Fiala, Andr{\'e} and Tavosanis, Gaia},
	citation = {eLife 2021;10:e74172},
	doi = {10.7554/eLife.74172},
	editor = {Ramaswami, Mani and VijayRaghavan, K},
	issn = {2050-084X},
	journal = {eLife},
	month = {dec},
	pages = {e74172},
	pub_date = {2021-12-29},
	publisher = {eLife Sciences Publications, Ltd},
	title = {The anterior paired lateral neuron normalizes odour-evoked activity in the \textit{Drosophila} mushroom body calyx},
	url = {https://doi.org/10.7554/eLife.74172},
	volume = 10,
	year = 2021}

@article{CKH24,
	author = {Choi, Kiri and Kim, Won Kyu and Hyeon, Changbong},
	doi = {10.1103/PhysRevResearch.6.023298},
	issue = {2},
	journal = {Phys. Rev. Res.},
	month = {Jun},
	numpages = {19},
	pages = {023298},
	publisher = {American Physical Society},
	title = {Unveiling the odor representation in the inner brain of Drosophila through compressed sensing},
	url = {https://link.aps.org/doi/10.1103/PhysRevResearch.6.023298},
	volume = {6},
	year = {2024}}

@article{BDW23,
	author = {Barth-Maron, Asa and D'Alessandro, Isabel and Wilson, Rachel I.},
	date = {2023/12/04},
	doi = {10.1016/j.cub.2023.10.041},
	isbn = {0960-9822},
	journal = {Current Biology},
	journal1 = {Current Biology},
	month = {2026/08/10},
	n2 = {Gain control is a process that adjusts a system?s sensitivity when input levels change. Neural systems contain multiple mechanisms of gain control, but we do not understand why so many mechanisms are needed or how they interact. Here, we investigate these questions in the Drosophila antennal lobe, where we identify several types of inhibitory interneurons with specialized gain control functions. We find that some interneurons are nonspiking, with compartmentalized calcium signals, and they specialize in intra-glomerular gain control. Conversely, we find that other interneurons are recruited by strong and widespread network input; they specialize in global presynaptic gain control. Using computational modeling and optogenetic perturbations, we show how these mechanisms can work together to improve stimulus discrimination while also minimizing temporal distortions in network activity. Our results demonstrate how the robustness of neural network function can be increased by interactions among diverse and specialized mechanisms of gain control.},
	number = {23},
	pages = {5109--5120.e7},
	publisher = {Elsevier},
	title = {Interactions between specialized gain control mechanisms in olfactory processing},
	type = {doi: 10.1016/j.cub.2023.10.041},
	url = {https://doi.org/10.1016/j.cub.2023.10.041},
	volume = {33},
	year = {2023},
	year1 = {2023}}

@article{LLM20,
	article_type = {journal},
	author = {Li, Feng and Lindsey, Jack W and Marin, Elizabeth C and Otto, Nils and Dreher, Marisa and Dempsey, Georgia and Stark, Ildiko and Bates, Alexander S and Pleijzier, Markus William and Schlegel, Philipp and Nern, Aljoscha and Takemura, Shin-ya and Eckstein, Nils and Yang, Tansy and Francis, Audrey and Braun, Amalia and Parekh, Ruchi and Costa, Marta and Scheffer, Louis K and Aso, Yoshinori and Jefferis, Gregory SXE and Abbott, Larry F and Litwin-Kumar, Ashok and Waddell, Scott and Rubin, Gerald M},
	citation = {eLife 2020;9:e62576},
	doi = {10.7554/eLife.62576},
	editor = {Griffith, Leslie C and Marder, Eve and Griffith, Leslie C and Pipkin, Jason and Doe, Chris Q},
	issn = {2050-084X},
	journal = {eLife},
	month = {dec},
	pages = {e62576},
	pub_date = {2020-12-14},
	publisher = {eLife Sciences Publications, Ltd},
	title = {The connectome of the adult Drosophila mushroom body provides insights into function},
	url = {https://doi.org/10.7554/eLife.62576},
	volume = 9,
	year = 2020}

@article{LTZ22,
	author = {Lazar, Aurel A. and Turkcan, Mehmet Kerem and Zhou, Yiyin},
	doi = {10.3389/fninf.2022.853098},
	issn = {1662-5196},
	journal = {Frontiers in Neuroinformatics},
	title = {A Programmable Ontology Encompassing the Functional Logic of the Drosophila Brain},
	url = {https://www.frontiersin.org/journals/neuroinformatics/articles/10.3389/fninf.2022.853098},
	volume = {Volume 16 - 2022},
	year = {2022}}

@article{CAD05,
	author = {Couto, Africa and Alenius, Mattias and Dickson, Barry J.},
	date = {2005/09/06},
	doi = {10.1016/j.cub.2005.07.034},
	isbn = {0960-9822},
	journal = {Current Biology},
	journal1 = {Current Biology},
	month = {2026/05/01},
	n2 = {Background: Olfactory receptor neurons (ORNs) convey chemical information into the brain, producing internal representations of odors detected in the periphery. A comprehensive understanding of the molecular and neural mechanisms of odor detection and processing requires complete maps of odorant receptor (Or) expression and ORN connectivity, preferably at single-cell resolution.Results: We have constructed near-complete maps of Or expression and ORN targeting in the Drosophila olfactory system. These maps confirm the general validity of the ?one neuron?one receptor? and ?one glomerulus?one receptor? principles and reveal several additional features of olfactory organization. ORNs in distinct sensilla types project to distinct regions of the antennal lobe, but neighbor relations are not preserved. ORNs grouped in the same sensilla do not express similar receptors, but similar receptors tend to map to closely appositioned glomeruli in the antennal lobe. This organization may serve to ensure that odor representations are dispersed in the periphery but clustered centrally. Integrated with electrophysiological data, these maps also predict glomerular representations of specific odorants. Representations of aliphatic and aromatic compounds are spatially segregated, with those of aliphatic compounds arranged topographically according to carbon chain length.Conclusions: These Or expression and ORN connectivity maps provide further insight into the molecular, anatomical, and functional organization of the Drosophila olfactory system. Our maps also provide an essential resource for investigating how internal odor representations are generated and how they are further processed and transmitted to higher brain centers.},
	number = {17},
	pages = {1535--1547},
	publisher = {Elsevier},
	title = {Molecular, Anatomical, and Functional Organization of the \textit{Drosophila} Olfactory System},
	type = {doi: 10.1016/j.cub.2005.07.034},
	url = {https://doi.org/10.1016/j.cub.2005.07.034},
	volume = {15},
	year = {2005},
	year1 = {2005}}

@inbook{LV08,
	address = {New York, NY},
	author = {Laissue, Philippe P. and Vosshall, Leslie B.},
	booktitle = {Brain Development in Drosophila melanogaster},
	doi = {10.1007/978-0-387-78261-4_7},
	editor = {Technau, Gerhard M.},
	isbn = {978-0-387-78261-4},
	pages = {102--114},
	publisher = {Springer New York},
	title = {The Olfactory Sensory Map in Drosophila},
	url = {https://doi.org/10.1007/978-0-387-78261-4_7},
	year = {2008}}

@article{LAY18,
	author = {Lazar, Aurel A. and Yeh, Chung-Heng},
	journal = {BMC Neuroscience},
	number = {Suppl 2},
	pages = {F3},
	title = {A Molecular Odorant Transduction Model and Combinatorial Encoding in the Drosophila Antennae},
	volume = {19},
	year = 2018}

@article{LLYZ24,
	author = {Lazar, Aurel A. and Liu, Tingkai and Yeh, Chung-Heng and Zhou, Yiyin},
	doi = {10.3389/fphys.2024.1410946},
	issn = {1664-042X},
	journal = {Frontiers in Physiology},
	title = {Modeling and characterization of pure and odorant mixture processing in the Drosophila mushroom body calyx},
	url = {https://www.frontiersin.org/journals/physiology/articles/10.3389/fphys.2024.1410946},
	volume = {Volume 15 - 2024},
	year = {2024}}

@article{LLY23,
	author = {Lazar, Aurel A. and Liu, Tingkai and Yeh, Chung-Heng},
	doi = {10.1371/journal.pcbi.1011043},
	journal = {PLOS Computational Biology},
	month = {04},
	note = {doi: 10.1371/journal.pcbi.1011043},
	number = {4},
	pages = {1-33},
	publisher = {Public Library of Science},
	title = {The functional logic of odor information processing in the Drosophila antennal lobe},
	url = {https://doi.org/10.1371/journal.pcbi.1011043},
	volume = {19},
	year = {2023}}

@article{KLS15,
	author = {Kim, Anmo J. and Lazar, Aurel A. and Slutskiy, Yevgeniy B.},
	doi = {10.7554/eLife.06651},
	journal = {eLife},
	pages = {e06651},
	publisher = {eLife Sciences Publications Limited},
	title = {Projection neurons in Drosophila antennal lobes signal the acceleration of odor concentrations},
	year = {2015}}

@article{LAY20,
	author = {Lazar, Aurel A. and Yeh, Chung-Heng},
	doi = {10.1371/journal.pcbi.1007751},
	editor = {Louis, MatthieuEditor},
	issn = {1553-7358},
	journal = {PLOS Computational Biology},
	month = {4},
	number = {4},
	pages = {e1007751},
	publisher = {Public Library of Science (PLoS)},
	title = {A molecular odorant transduction model and the complexity of spatio-temporal encoding in the {D}rosophila antenna},
	volume = {16},
	year = {2020}}

@article{CCH11,
	author = {Ann-Shyn Chiang and Chih-Yung Lin and Chao-Chun Chuang and Hsiu-Ming Chang and Chang-Huain Hsieh and Chang-Wei Yeh and Chi-Tin Shih and Jian-Jheng Wu and Guo-Tzau Wang and Yung-Chang Chen and Cheng-Chi Wu and Guan-Yu Chen and Yu-Tai Ching and Ping-Chang Lee and Chih-Yang Lin and Hui-Hao Lin and Chia-Chou Wu and Hao-Wei Hsu and Yun-Ann Huang and Jing-Yi Chen and Hsin-Jung Chiang and Chun-Fang Lu and Ru-Fen Ni and Chao-Yuan Yeh and Jenn-Kang Hwang},
	doi = {https://doi.org/10.1016/j.cub.2010.11.056},
	issn = {0960-9822},
	journal = {Current Biology},
	number = {1},
	pages = {1-11},
	title = {Three-Dimensional Reconstruction of Brain-wide Wiring Networks in Drosophila at Single-Cell Resolution},
	url = {https://www.sciencedirect.com/science/article/pii/S0960982210015228},
	volume = {21},
	year = {2011}}

@article{SRW10,
	author = {Seki, Yoichi and Rybak, J\"{u}rgen and Wicher, Dieter and Sachse, Silke and Hansson, Bill S.},
	doi = {10.1152/jn.00249.2010},
	eprint = {https://doi.org/10.1152/jn.00249.2010},
	journal = {Journal of Neurophysiology},
	note = {PMID: 20505124},
	number = {2},
	pages = {1007-1019},
	title = {Physiological and Morphological Characterization of Local Interneurons in the Drosophila Antennal Lobe},
	url = {https://doi.org/10.1152/jn.00249.2010},
	volume = {104},
	year = {2010}}

@article{TEI12,
	author = {Tanaka, Nobuaki K. and Endo, Keita and Ito, Kei},
	doi = {https://doi.org/10.1002/cne.23142},
	eprint = {https://onlinelibrary.wiley.com/doi/pdf/10.1002/cne.23142},
	journal = {Journal of Comparative Neurology},
	number = {18},
	pages = {4067-4130},
	title = {Organization of antennal lobe-associated neurons in adult Drosophila melanogaster brain},
	url = {https://onlinelibrary.wiley.com/doi/abs/10.1002/cne.23142},
	volume = {520},
	year = {2012}}

@article{RVM11,
	author = {Rivera-Alba, Marta and Vitaladevuni, Shiv N and Mishchenko, Yuriy and Lu, Zhiyuan and Takemura, Shin-Ya and Scheffer, Lou and Meinertzhagen, Ian A and Chklovskii, Dmitri B and de Polavieja, Gonzalo G},
	journal = {Current Biology},
	month = dec,
	number = 23,
	pages = {2000--2005},
	publisher = {Elsevier},
	title = {Wiring Economy and Volume Exclusion Determine Neuronal Placement in the Drosophila Brain},
	volume = 21,
	year = 2011}

@article{TBL13,
	author = {Takemura, Shin-Ya and Bharioke, Arjun and Lu, Zhiyuan and Nern, Aljoscha and Vitaladevuni, Shiv and Rivlin, Patricia K and Katz, William T and Olbris, Donald J and Plaza, Stephen M and Winston, Philip and Zhao, Ting and Horne, Jane Anne and Fetter, Richard D and Takemura, Satoko and Blazek, Katerina and Chang, Lei-Ann and Ogundeyi, Omotara and Saunders, Mathew A and Shapiro, Victor and Sigmund, Christopher and Rubin, Gerald M and Scheffer, Louis K and Meinertzhagen, Ian A and Chklovskii, Dmitri B},
	journal = {Nature},
	month = aug,
	number = 7461,
	pages = {175--181},
	title = {A visual motion detection circuit suggested by Drosophila connectomics},
	volume = 500,
	year = 2013}

@article{AHY14,
	article_type = {journal},
	author = {Aso, Yoshinori and Hattori, Daisuke and Yu, Yang and Johnston, Rebecca M and Iyer, Nirmala A and Ngo, Teri-TB and Dionne, Heather and Abbott, LF and Axel, Richard and Tanimoto, Hiromu and Rubin, Gerald M},
	citation = {eLife 2014;3:e04577},
	doi = {10.7554/eLife.04577},
	editor = {Griffith, Leslie C},
	issn = {2050-084X},
	journal = {eLife},
	month = {dec},
	pages = {e04577},
	pub_date = {2014-12-23},
	publisher = {eLife Sciences Publications, Ltd},
	title = {The neuronal architecture of the mushroom body provides a logic for associative learning},
	url = {https://doi.org/10.7554/eLife.04577},
	volume = 3,
	year = 2014}

@article{SXJ20,
	article_type = {journal},
	author = {Scheffer, Louis K and Xu, C Shan and Januszewski, Michal and Lu, Zhiyuan and Takemura, Shin-ya and Hayworth, Kenneth J and Huang, Gary B and Shinomiya, Kazunori and Maitlin-Shepard, Jeremy and Berg, Stuart and Clements, Jody and Hubbard, Philip M and Katz, William T and Umayam, Lowell and Zhao, Ting and Ackerman, David and Blakely, Tim and Bogovic, John and Dolafi, Tom and Kainmueller, Dagmar and Kawase, Takashi and Khairy, Khaled A and Leavitt, Laramie and Li, Peter H and Lindsey, Larry and Neubarth, Nicole and Olbris, Donald J and Otsuna, Hideo and Trautman, Eric T and Ito, Masayoshi and Bates, Alexander S and Goldammer, Jens and Wolff, Tanya and Svirskas, Robert and Schlegel, Philipp and Neace, Erika and Knecht, Christopher J and Alvarado, Chelsea X and Bailey, Dennis A and Ballinger, Samantha and Borycz, Jolanta A and Canino, Brandon S and Cheatham, Natasha and Cook, Michael and Dreher, Marisa and Duclos, Octave and Eubanks, Bryon and Fairbanks, Kelli and Finley, Samantha and Forknall, Nora and Francis, Audrey and Hopkins, Gary Patrick and Joyce, Emily M and Kim, SungJin and Kirk, Nicole A and Kovalyak, Julie and Lauchie, Shirley A and Lohff, Alanna and Maldonado, Charli and Manley, Emily A and McLin, Sari and Mooney, Caroline and Ndama, Miatta and Ogundeyi, Omotara and Okeoma, Nneoma and Ordish, Christopher and Padilla, Nicholas and Patrick, Christopher M and Paterson, Tyler and Phillips, Elliott E and Phillips, Emily M and Rampally, Neha and Ribeiro, Caitlin and Robertson, Madelaine K and Rymer, Jon Thomson and Ryan, Sean M and Sammons, Megan and Scott, Anne K and Scott, Ashley L and Shinomiya, Aya and Smith, Claire and Smith, Kelsey and Smith, Natalie L and Sobeski, Margaret A and Suleiman, Alia and Swift, Jackie and Takemura, Satoko and Talebi, Iris and Tarnogorska, Dorota and Tenshaw, Emily and Tokhi, Temour and Walsh, John J and Yang, Tansy and Horne, Jane Anne and Li, Feng and Parekh, Ruchi and Rivlin, Patricia K and Jayaraman, Vivek and Costa, Marta and Jefferis, Gregory SXE and Ito, Kei and Saalfeld, Stephan and George, Reed and Meinertzhagen, Ian A and Rubin, Gerald M and Hess, Harald F and Jain, Viren and Plaza, Stephen M},
	citation = {eLife 2020;9:e57443},
	doi = {10.7554/eLife.57443},
	editor = {Marder, Eve and Eisen, Michael B and Pipkin, Jason and Doe, Chris Q},
	issn = {2050-084X},
	journal = {eLife},
	month = {sep},
	pages = {e57443},
	pub_date = {2020-09-03},
	publisher = {eLife Sciences Publications, Ltd},
	title = {A connectome and analysis of the adult \textit{Drosophila} central brain},
	url = {https://doi.org/10.7554/eLife.57443},
	volume = 9,
	year = 2020}

@article{SBS21,
	article_type = {journal},
	author = {Schlegel, Philipp and Bates, Alexander Shakeel and St{\"u}rner, Tomke and Jagannathan, Sridhar R and Drummond, Nikolas and Hsu, Joseph and Serratosa Capdevila, Laia and Javier, Alexandre and Marin, Elizabeth C and Barth-Maron, Asa and Tamimi, Imaan FM and Li, Feng and Rubin, Gerald M and Plaza, Stephen M and Costa, Marta and Jefferis, Gregory S X E},
	citation = {eLife 2021;10:e66018},
	doi = {10.7554/eLife.66018},
	editor = {Griffith, Leslie C and Dulac, Catherine and Luo, Liqun},
	issn = {2050-084X},
	journal = {eLife},
	month = {may},
	pages = {e66018},
	pub_date = {2021-05-25},
	publisher = {eLife Sciences Publications, Ltd},
	title = {Information flow, cell types and stereotypy in a full olfactory connectome},
	url = {https://doi.org/10.7554/eLife.66018},
	volume = 10,
	year = 2021}

@article{OSF15,
	author = {Ohyama, Tomoko and Schneider-Mizell, Casey M and Fetter, Richard D and Aleman, Javier Valdes and Franconville, Romain and Rivera-Alba, Marta and Mensh, Brett D and Branson, Kristin M and Simpson, Julie H and Truman, James W and Cardona, Albert and Zlatic, Marta},
	journal = {Nature},
	month = apr,
	number = 7549,
	pages = {633--639},
	title = {A multilevel multimodal circuit enhances action selection in Drosophila},
	volume = 520,
	year = 2015}

\end{document}